\documentclass[useAMS,twocolumn]{emulateapj}
\usepackage[latin2]{inputenc}

\def\apjl{{ApJ}}                
\def\apjs{{ApJS}}               
\def\ao{{Appl.~Opt.}}           
\def\aap{{A\&A}}                

\usepackage{enumerate}

\newcommand{\mpc}{\rm {h^{-1}Mpc }}

\begin{document}

\title{Cosmological Density Fluctuations on 100Mpc Scales and their ISW Effect}

\author{P\'eter P\'apai\altaffilmark{1,2} and Istv\'an Szapudi\altaffilmark{2}}
\altaffiltext{1}{Department of Physics and Astronomy, University of Hawaii, 2505 Correa Road, HI, USA}
\altaffiltext{2}{Institute for Astronomy, University of Hawaii, 2680 Woodlawn Drive, HI, USA}

\begin{abstract}

We measure the matter probability distribution function (PDF) via counts in cells in a volume limited subsample of the Sloan Digital Sky Survey Luminous Red Galaxy Catalog on scales from $30~h^{-1}$Mpc to $150~h^{-1}$Mpc and estimate the linear Integrated Sachs--Wolfe effect produced by supervoids and superclusters in the tail of the PDF.

We characterize the PDF by the variance, $S_3$, and $S_4$, and study in simulations the systematic effects due to finite volume, survey shape and redshift distortion. We compare our measurement to the prediction of $\Lambda$CDM with linear bias and find a good agreement. 

We use the moments to approximate the tail of the PDF with analytic functions. A simple Gaussian model for the superstructures appears to be consistent with the claim by \citeauthor{Granett2008} that density fluctuations on $100~h^{-1}$Mpc scales produce hot and cold spots with $\Delta T \approx 10\mu K$ on the cosmic microwave background.

\end{abstract}


\section{Introduction}

After last scattering photons traveled through mostly neutral media. Although radiation and matter are not strongly coupled, there is still a secondary signal due to large scale structure on top of the primary fluctuations of the cosmic microwave background (CMB) radiation. The Integrated Sachs-Wolfe (ISW) effect \citep{SW1967} accounts for most of the secondary anisotropies for low multipoles \citep{Hu2002}. As the expansion of the universe accelerates, gravitational potential wells and hills decay. Photons traversing these get blueshifted or redshifted. 

Due to its weak signal, ISW detection is very challenging. Cross-correlating galaxy surveys with CMB maps yield results from marginally significant \citep{Scranton2003,Afshordi2004,Padmanabhan2005,Raccanelli2008,Sawangwit2010} to $4.5\sigma$ detections \citep{Giannantonio2008,Ho2008}. The higher significance was achieved by a joint analysis of surveys. Other techniques focusing on the signal from discrete objects can reach up to $4.5\sigma$ from a single survey \citep{McEwen2008,Granett2008}. 

The ISW effect can be a unique probe of dark energy if well-measured. From cross-correlation measurements and the \emph{Wilkinson Microwave Anisotropy Probe} (\emph{WMAP}) power spectrum \citep{Bennett2003} it has already been shown that it is possible to constrain cosmological parameters \citep{Giannantonio2008,Ho2008}. Despite the fact that the detection of the signal from discrete objects has higher significance, they cannot be used for parameter estimation due to the lack of simple quantitative models. 

Further motivation for studying super structures stems from anomalies in the low l modes of the CMB \citep{Tegmark2003,Copi2004}. \cite{Inoue2007} calculate the effect of large, dust filled, compensated voids in the local universe. They were successful in explaining the observed CMB anomalies but these voids, due to their size and depth, do not fit into the widely accepted picture of clustering. They assume extra power on large scales in the matter power spectrum. When subtracting the estimated local ISW signal from CMB maps, \cite{Francis2009} found that the significance of the anomalies decreased.

In this paper, our principal goal is to give an estimate of the ISW signal coming from large overdense or underdense regions (superclusters or supervoids). To achieve this we have to deal with two separate problems: the density of the extreme fluctuations and the ISW effect associated with them. In Section \ref{sM}, we present our measurement of counts in cells (CIC) in the Sloan Digital Sky Server (SDSS) Luminous Red Galaxy (LRG) Catalog. From CIC we derive the first few moments (the variance, $S_3$, and $S_4$) of the matter probability distribution function (PDF) and compare them to their theoretical values. In Section \ref{sI}, we use these to estimate the PDF focusing on its tail, since an enhanced tail could explain a strong ISW signal from super structures. We use a simple Gaussian model to derive an expression for their profile and the potential. We compare their estimated ISW signal to the elusive results of \cite{Granett2008}. More about their ISW measurement can be found in Section \ref{sI} of this paper. In Section \ref{dis}, we summarize and discuss our results.

\section{Measurements}

\label{sM}
The hydrodynamical model of the universe is based on the assumption that the observed galaxy distribution is a Poisson-sampled version of a continuous field. Furthermore, this continuous field is a realization of a random field (see, e.g., \citealt{Peebles1980}). 
Measuring CIC is a well-established method to estimate its PDF (see, e.g., \citealt{Colombi2000,Szapudi2000}). In this Section first we give a brief summary of this method and description of the data and the algorithm we use in our analysis. Then we compare our findings to the predictions of $\Lambda$CDM.

   \subsection{CIC and the PDF}
   
In the following we deal with two random fields, one corresponding to dark matter, and the other its biased version, the galaxy field. In our notation the terms that refer to these are "matter PDF", "matter field", "galaxy PDF" and "galaxy field". 

The matter PDF for cells is fully given by its cumulants, possibly normalized \citep{Peebles1980}:
\begin{eqnarray}
&\overline{\xi}_n = \frac{1}{V^n} \int_{V} ... \int_{V} \big<\delta (x_1)...\delta (x_n)\big>_{c}dx_1 ... dx_n ,&\\
&S_{n} = \frac{\overline{\xi}_n}{\overline{\xi}^{n-1}_2},&
\end{eqnarray}
where $V$ is the volume of a cell and the subscript $c$ refers to connected moments. 
When $\overline{\xi}_2 << 1$, $S_n$'s are around unity according to perturbation theory. Thus, $S_3$, the skewness, and $S_4$, the kurtosis represent the lowest order correction to a Gaussian distribution. Deriving these quantities from galaxy counts can be done in two steps: first, by using factorial moments to get the moments of the underlying galaxy PDF (see e.g. \citealt{Szapudi1993}), second, by using some galaxy-dark matter biasing scheme to transform the galaxy PDF into the matter PDF. 

Estimators for the variance ($\overline{\xi}_2$), $S_3$ and $S_4$ are:
\begin{eqnarray}
&\overline{\xi}_2 = \frac{\big<N^2\big>-\big<N\big>}{\big<N\big>^2}-1, \label{var}&\\
&S_3 = \frac{(F_3-3F_2F_1+2F_1^3)/F_1^3}{\overline{\xi}_2^2}, \label{s3}&\\
&S_4 = \frac{(F_4-4F_3F_1+12F_2F_1^2-6F_1^4-3F_2^2)/F_1^4}{\overline{\xi}_2^3}. \label{s4}&
\end{eqnarray}
where
\begin{eqnarray}
&F_n = \big<N(N-1)...(N-n+1)\big>.&
\end{eqnarray}
Ergodicity ensures that these ensemble averages can be calculated from a single, ideally large, volume limited survey. The caveats, arising when we depart from the ideal case, are discussed later in this Section. 

In this paper we use cosmological parameters taken at their best-fit \emph{WMAP} values \citep{Spergel2007}. For the bias we fit the simplest, deterministic, local, linear model:
\begin{eqnarray}
&\delta_g = b \delta .\label{bias}&
\end{eqnarray}
This is generally a good approximation on quasi-linear scales. Its validity is tested in Section \ref{ssMeas}.

   \subsection{The Data}

\label{data}

Among the spectroscopic galaxy surveys available today the Seventh Data Release of the Sloan Digital Sky Survey (SDSS DR7) covers the largest volume \citep{Abazajian2009}. The LRG sample is generally regarded as a good cosmological probe. Properties of the LRGs can be found in \cite{Eisenstein2001}. In practice, flags in the Sloan database identify these galaxies. For CIC one needs a volume limited sample which can be obtained by magnitude and redshift cuts. We restricted our LRG sample to redshifts between 0.24 and 0.31 and $k$-corrected absolute magnitudes between $-$22.3 and $-$24.3 in the $r$ band. We used values from the Photoz and SpecObjAll tables on the SkyServer Web site. By excluding the three stripes in the Southern Galactic Cap we were left with 21613 galaxies. After converting redshift into comoving radial distance by using the best \emph{WMAP} cosmological parameters, the angle-averaged density appears to be uniform with fluctuations consistent with Poisson noise. The selection function of a similar data set is plotted in Figure 12 in \cite{Eisenstein2001}.

   \subsection{The Algorithm}

\label{ssAlg}
From R.A., decl., and $z$ coordinates, we calculated comoving Cartesian coordinates. Then we placed a rectangular grid over the sample. Since in this arrangement cubical cells are readily accessible, we chose to measure CIC in cubes. The survey mask, however, has a complex shape, usually given by spherical polygons. Holes and the irregular boundary cause unwanted edge effects which could bias the results in a complicated way. 

To tackle this problem, first we took a cube-shaped region encompassing the survey. Then we created two negatives by filling the parts in the mask and outside the survey area with dummy galaxies from a Poisson point process; one with the average density of the survey and one with hundred times that density. We added the first negative to the survey to fill the holes. We measured CIC in this and in the second negative in parallel. Since its density is large, the counts from the second negative provide a good measure of the overlap of the cells with the survey geometry.
We ignored any cell that had more than 10\% of its volume outside, which corresponded to having a galaxy count in the second negative larger than 100 X average density X volume of cell X 0.1. In this work we used MANGLE \citep{Swanson2008} to check whether an object was inside the mask.

   \subsection{The Systematics}
   
\label{sys}      
The shape of the survey and diluting the data with a Poisson point process introduce systematic bias into our measurements of CIC. In order to assess its level we studied simulations. We created mock catalogs with a second-order Lagrangian (2LPT) code \citep{Crocce2006}. We created 100 mock catalogs in $2500~h^{-1}$Mpc cubes, then we used these to create another set of mocks by applying the mask of the spectroscopic survey. The galaxies were downsampled in every case to match the average density with that of the data. With these and the two negatives described in Section \ref{ssAlg}, we were able to measure CIC in three different arrangements.

\begin{enumerate}[(i)]

\item Ideal, large, cubic-shaped simulations. \label{s1}

\item  Variant of (\ref{s1}). We only took into account a cell when at least 90\% of its volume lay inside the survey area. \label{s2}

\item  Dropping galaxies outside the survey area and filling it with the dummy galaxies of the negative to preserve the average density. (The 90\% rule still applies) \label{s3}

\end{enumerate}

\begin{figure}	
        \begin{center}
        \includegraphics[scale=0.45]{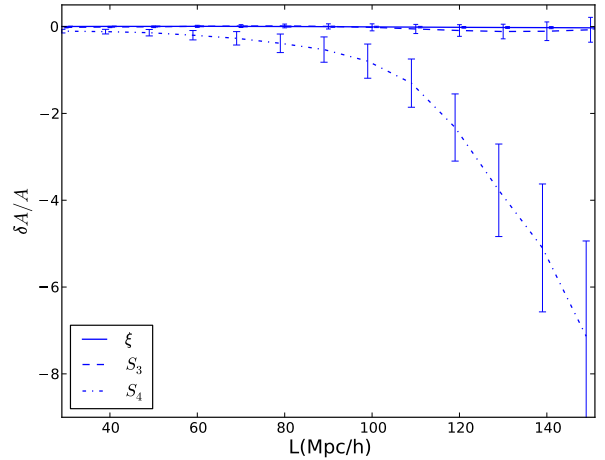}
        \end{center}
        \caption{Bias due to the finite volume of the survey as a function of the cell size. \label{fig1}}
\end{figure}

First we estimated the cosmic bias due to the survey shape and volume. This question has been studied extensively in the past (see, e.g., \citealt{Szapudi1996}). We measured the variance, $S_3$ and $S_4$ in the first two arrangements (\ref{s1}) and (\ref{s2}). In Figure \ref{fig1} the ratio, $\delta A/A$, is plotted for each of these quantities, where $A$ is the quantity measured in arrangement (\ref{s1}) . Error bars were estimated from the scatter around the average. The error of the average is plotted, so the error of a single measurement is ten times larger. In the case of the variance this is the well-understood integral constraint problem and the ratio does not exceed a couple of percents even at the largest scale. For $S_3$ the ratio is consistent with 1 but for $S_4$ the difference from 1 is not negligible even at relatively small scales. However, as we show later, this bias is still small compared to the cosmic error.

The bias, caused by the data having been diluted with a random sample, can be understood in case that evenly distributed holes comprise the mask. The resulting catalog can be considered the linear combination of the Poisson sampled galaxy and a constant field. This constant field is the random sample that fills the mask:
\begin{eqnarray}
&\rho = \rho_{galaxy} + \rho_{random}.&
\end{eqnarray}
Subsequently the density contrast can be written as
\begin{eqnarray}
&\delta = y\delta_{galaxy} + (1-y)\delta_{random}&
\end{eqnarray}
with
\begin{eqnarray}
&y = \frac{\overline{\rho}_{galaxy}}{\overline{\rho}} \label{y0}.&
\end{eqnarray}
 
Since $\delta_{random}$ is zero the moments of $\delta$ are proportional to the moments of $\delta_{data}$ :
\begin{eqnarray}
&\overline{\xi} = y^2 \overline{\xi}_{galaxy} , \label{y1}&\\
&S_n = y^{-n+2}  S^{galaxy}_n \label{y2}.&
\end{eqnarray}

\begin{figure}	
        \begin{center}
        \includegraphics[scale=0.45]{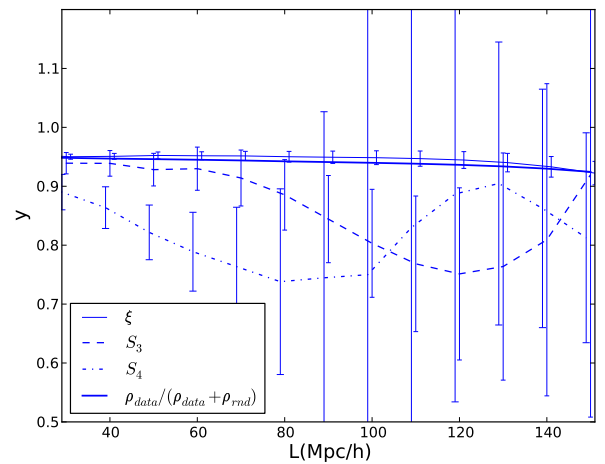}
        \end{center}
        \caption{$y$ as a function of cell size from Equation (\ref{y1}) and (\ref{y2}). \label{fig2}}
\end{figure}

We used arrangement (\ref{s2}) and (\ref{s3}) to express $y$ according to Equation (\ref{y1}) and Equation (\ref{y2}) (Figure \ref{fig2}). The measured values are consistent with our assumption, Equation (\ref{y0}). The robustness of this simple model is due to the fact that the variance and $S_n$ are insensitive to small changes in the cell shape  (e.g., in the case of a power-law correlation function the $S_n$s are constants, see \citealt{Peebles1980} or \citealt{Boschan1994,Szapudi1998} for a study of on this). This bias can be corrected for by measuring $y$ directly. 

\begin{figure}	
        \begin{center}
        \includegraphics[scale=0.45]{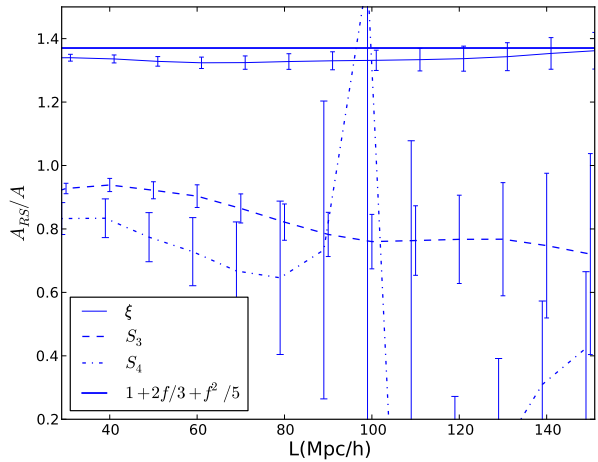}
        \end{center}
        \caption{Effect of redshift distortion on the variance ($\xi$), $S_3$, and $S_4$ as a function of the cell size. \label{fig3}}
\end{figure}

As the next step we added redshift distortions to the mock catalogs. The effect on the variance is expected to be similar to the effect on the monopole of the two-point function. $S_3$ and $S_4$ are affected less according to, e.g., \cite{Hivon1995}. While the Kaiser formula \citep{Kaiser1987,Hamilton1992} gives a good  description of this in the linear regime. The lowest order of the three-point function in Fourier space has been worked out by \cite{Scoccimarro1999} but Fourier transforming it back to redshift space is infeasible in general. Higher moments are gradually harder to compute. For these reasons we follow a phenomenological approach. In Figure \ref{fig3}, the ratios of redshift and real space values of the variance, $S_3$ and $S_4$ are plotted. The thick line is the predicted amplification from the Kaiser formula for the variance:
\begin{eqnarray}
&\overline{\xi}_{RS} = (1+2f/3+f^2/5) \overline{\xi},\label{KF}&
\end{eqnarray}
with
\begin{eqnarray}
&f = \Omega^{0.6}/b ,&
\end{eqnarray}
where the subscript RS stands for redshift space. In the simulations $b$ is 1. In this paper we assume that the effect of redshift distortion is small compared to the cosmic errors in the case of $S_3$ and $S_4$, and that the variance is amplified according to Equation (\ref{KF}).

\begin{figure}	
        \begin{center}
        \includegraphics[scale=0.45]{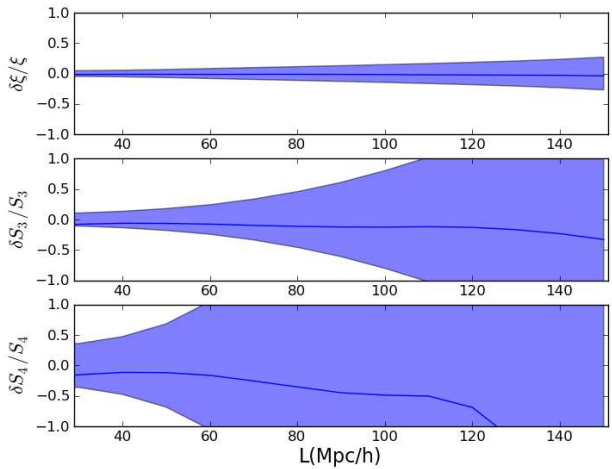}
        \end{center}
\caption{Total expected systematic bias is plotted after corrections as discussed in Section \ref{sys}. The shaded region represents the cosmic error.\label{fig4}}
\end{figure}

In practice if the systematic bias is small compared to the cosmic error then it is negligible. In Figure \ref{fig4} the total systematic bias after corrections according to Equation (\ref{y1}) and (\ref{y2}) and the cosmic error are plotted. It can be concluded that the proposed corrections are sufficient to measure the variance, $S_3$, and $S_4$ with an error that is not significantly different from the cosmic error. This plot also tells us that the signal-to-noise ratio drops below 1 around $100~h^{-1}$Mpc for $S_3$ and around $50~h^{-1}$Mpc for $S_4$, so they cannot be measured reliably beyond these scales. 

   \subsection{The Variance, $S_3$ and $S_4$}

\label{ssMeas}
We measured CIC in the SDSS DR7 spectroscopic LRG sample and compare the prediction of $\Lambda$CDM to our results.

For measuring the variance, $S_3$ and $S_4$ we followed the procedure outlined in Section \ref{ssAlg} and we corrected for the systematic bias as given by Equation (\ref{y1}) and (\ref{y2}). We determined the $y$ parameter from the simulations. 

Additionally, in the case of the real data one has to assume a galaxy--dark-matter biasing scheme. We used the simplest linear local model as given by Equation (\ref{bias}).
We computed the bias parameter from fitting the variance and testing its consistency on $S_3$ and $S_4$. We defined the following chi-square:
\begin{eqnarray}
&\chi^2(b) = (\overline{\xi}_d-\overline{\xi}_{th}(b))C^{-1}(\overline{\xi}_d-\overline{\xi}_{th}(b)).
\label{eq:}&
\end{eqnarray}
Here $\overline{\xi}_d$ and $\overline{\xi}_{th}$ stand for the measured and the theoretical variance. For the theory we used the real space linear model and we assumed that it transforms to redshift space as the monopole of the two-point function (see Equation (\ref{KF})). The measured variance was rescaled as in Equation (\ref{y1}) and extrapolated to present day ($z=0$) using the growth function (see, e.g., \citealt{Dodelson2003}).  The covariance matrix was calculated from mock catalogs described in Section \ref{sys}:  
\begin{eqnarray}
&C_{ij} = \frac{1}{N-1} \sum_{n}(\overline{\xi}_i^n-\overline{\overline{\xi}_i})(\overline{\xi}_j^n-\overline{\overline{\xi}_j}),&\\
&\overline{\overline{\xi}_j} = \frac{1}{N}\sum_{n}\overline{\xi}_j^n ,
\label{eqCov}&
\end{eqnarray}
where the superscript $n$ refers to the $n$th simulation and $N$ is the total number of simulations, in this case 100. These simulations are in redshift space and with bias equal to 1. After finding the minimum of the chi-square, the covariance matrix was rescaled according to Equation (\ref{KF}). 

\begin{figure}	
        \begin{center}
        \includegraphics[scale=0.45]{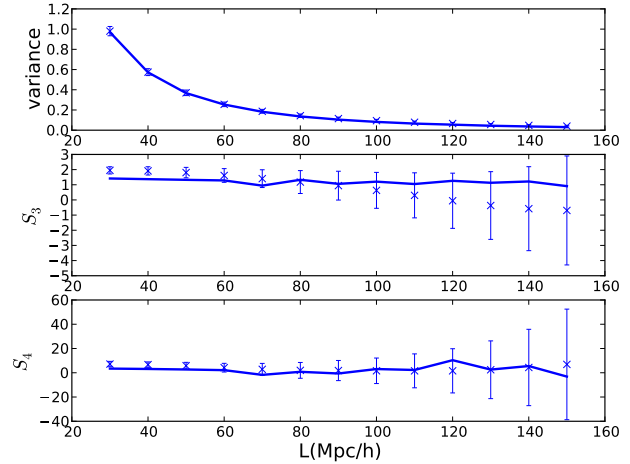}
        \end{center}
        \caption{Measured variance, $S_3$, and $S_4$ from the SDSS spectroscopic LRG subsample (see Section \ref{data} for a description of the data) with the theoretical predictions (solid line) vs. the cell size. \emph{L} is the size of a cubic cell.\label{fig6-8}}
\end{figure}

For our fit we used the range from $30~h^{-1}$Mpc to $150~h^{-1}$Mpc where the linear theory is generally assumed to be valid. The result is $b = 2.14^{+0.13}_{-0.14}$ with $1\sigma$ uncertainty, which is consistent with findings of \cite{Okumura2008}, who used a very similar data set. The best fitting variance is plotted with the data on the upper panel of Figure \ref{fig6-8}. When we changed the boundaries of the range to $50~h^{-1}$Mpc and $130~h^{-1}$Mpc, we found that the change in the best $b$ was consistent, only $+0.03$.  

We also measured $S_3$ and $S_4$ and applied Equation (\ref{y2}). In Figure \ref{fig6-8} these are plotted along with the prediction of linear $\Lambda$CDM \citep{Juszkiewicz1993,Bernardeau1994}:
\begin{eqnarray}
&bS_3 = \frac{34}{7} + \gamma_1 ,&\\
&b^2S_4 = \frac{60712}{1323} + \frac{62}{3}\gamma_1 + \frac{7}{3}\gamma_1^2 - \frac{2}{3}\gamma_2 &
\end{eqnarray}
with
\begin{eqnarray}
&\gamma_i = \frac{d \log^i \overline{\xi} }{d \log r^i}. &
\end{eqnarray}
The average of the correlation function, $\overline{\xi}$, was calculated via Monte Carlo simulations for cells of $30~h^{-1}$Mpc $+ \Delta,40~h^{-1}$Mpc $+ \Delta,..., 150~h^{-1}$Mpc $+ \Delta$, where $\Delta$ is $0$ or $\pm 3$. The $\gamma$ values were estimated using discrete derivatives of $\overline{\xi}$. The data values from the plot are collected in Table \ref{table1}. 

\begin{table}
\begin{center}
\begin{tabular}{|c|cccccc|}
\hline
$r(h^{-1}Mpc)$ & $\xi$ & $\Delta \xi$ & $S_3$ & $\Delta S_3$ & $S_4$ & $\Delta S_4$\\
\hline
30 & 0.980 & 0.046 & 1.951 & 0.238 & 7.176 & 2.225 \\
40 & 0.574 & 0.034 & 1.916 & 0.273 & 6.701 & 2.234 \\
50 & 0.370 & 0.027 & 1.804 & 0.343 & 5.821 & 2.770 \\
60 & 0.256 & 0.022 & 1.613 & 0.452 & 4.191 & 3.581 \\
70 & 0.187 & 0.019 & 1.404 & 0.586 &  &  \\
80 & 0.144 & 0.017 & 1.179 & 0.755 &  &  \\
90 & 0.115 & 0.015 & 0.942 & 0.951 &  &  \\
100 & 0.094 & 0.013 &  &  &  &  \\
110 & 0.078 & 0.011 &  &  &  & \\
120 & 0.066 & 0.010 &  &  &  &  \\
130 & 0.056 & 0.009 &  &  &  &  \\
140 & 0.048 & 0.008 &  &  &  &  \\
150 & 0.040 & 0.008 &  &  &  &  \\
\hline
\end{tabular}
\end{center}
\caption{The Numerical Values of $\xi$, $S_3$, and $S_4$ with $1\sigma$ Uncertainty. We leave the fields blank when the signal/noise ratio is less than one.}
\label{table1}
\end{table}

We tested the goodness of the theory by calculating a covariance matrix with the shrinkage technique \citep{Pope2008} with a diagonal of the empirical covariance matrix ($C$) as the target covariance matrix ($T$):
\begin{eqnarray}
&\tilde{C}_{ij} = \lambda T_{ij} + (1-\lambda)C_{ij}&
\end{eqnarray}
A recipe to calculate $\lambda$ is given in \cite{Pope2008}. This method ensures that we get a well behaving covariance matrix. We estimated the significance for the variance, $S_3$ and $S_4$ separately and jointly. The results are in Table \ref{table2}, showing a good overall agreement with our $\Lambda$CDM model with linear bias. Visually, the $S_n$'s appear to be slightly larger than expected on smaller scales, however, this is not statistically significant at all. On large scales, there appears to be a slight excess power on $130-150~h^{-1}$Mpc scales in the variance (not apparent in Figure \ref{fig6-8}). This is not large enough to influence the ISW effect, and its significance is only $2\sigma$ according to Table \ref{table2}. In summary,  $\Lambda$CDM is a good fit for all the moments we measured. Note that the shrinkage estimator ($\lambda = 0.0006$) gave identical results for the covariance matrix of the variance to that given by Equation (\ref{eqCov}).  

\begin{table}
\begin{center}
\begin{tabular}{|c|cc|}
\hline
 data & $ \lambda$ & p \\
\hline
$\xi$ & 0.0006 & 0.05  \\
$S_3$ & 0.13 & 0.80  \\
$S_4$ & 1. & 0.83  \\
Joint & 1. & 0.85  \\
\hline
\end{tabular}
\end{center}
\caption{Testing the Linear Model with Shrinkage Techique as in \cite{Pope2008}.}
\label{table2}
\end{table}

\section{The ISW Effect of Extreme Fluctuations}

\label{sI}
Here, we explore the possibility that the largest fluctuations in the linear matter density cause detectable anomalies on the CMB through the linear ISW effect. 

\cite{Granett2008} identified large underdense and overdense regions (supervoids and superclusters) in the SDSS DR4 photometric LRG sample. They stacked images, cut out from the CMB, centered on the directions of 50 supervoids and 50 superclusters found with the highest significance.

The signal is consistently present in every frequency band, so it is likely to have a cosmological origin. In this Section we put under scrutiny the tail of the density distribution and the density profile of supervoids or superclusters. The question we ask is: what is the expected ISW signal produced by the 50 objects with the highest and lowest densities in a survey similar to the one in \cite{Granett2008}? 

In order to answer this we have to have an estimate for tail of the matter PDF. We show that by using simple analytic functions to approximate the galaxy PDF this can be done robustly for our purposes. We also revise the way in which the ISW signal from these density extrema is estimated.

   \subsection{The Matter PDF}

If the galaxy PDF is known, a simple convolution with a Poisson-distribution gives the galaxy counts:
\begin{eqnarray}
&P(N) = \int{ \frac{\big<N\big>^N (1+\delta_g)^N}{N!}e^{-\big<N\big> (1+\delta_g)} P(\delta_g) d\delta_g} .
\label{eq:Poisson}&
\end{eqnarray}

\begin{figure*}	
    \begin{center}
        \includegraphics[scale=.6]{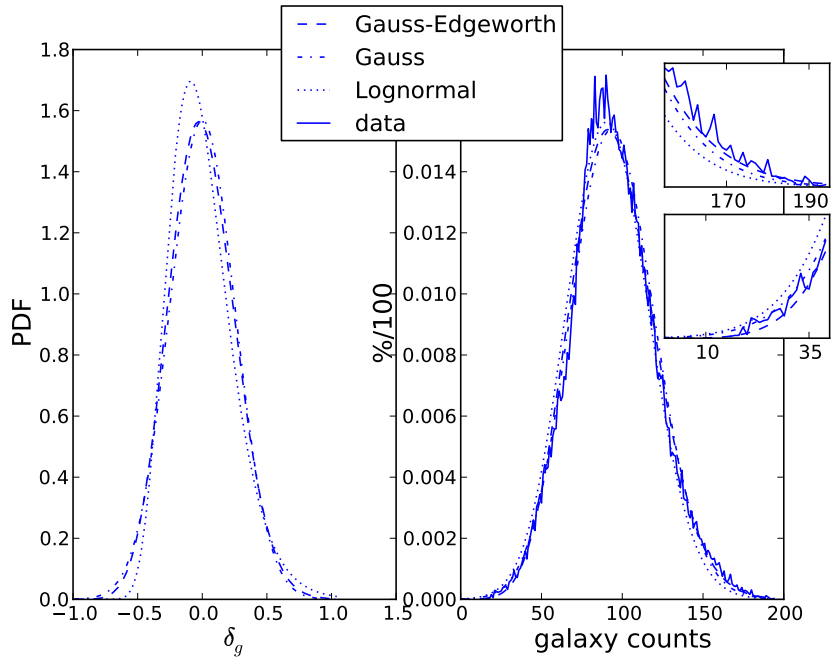}
    \end{center}
        \caption{On the left panel, three analytic functions approximating the galaxy PDF for cells of $100~h^{-1}$Mpc are plotted (for details, see the text). On the right panel, these are plotted again convolved with a Poisson noise (Equation (\ref{eq:Poisson})). The data is the solid line on the right panel.\label{fig9}}
\end{figure*}

On large scales one can approximate $P(\delta_g)$ as Gaussian, lognormal or second-order Edgeworth expansion (see, e.g., \citealt{Kim1998,Szapudi2004}). The first two depend on the variance only, while for the Edgeworth expansion we need $S_3$ and $S_4$ as well:
\begin{eqnarray}
&P_{G}(\delta_g) = \frac{1}{\sqrt{2\pi}\sigma}e^{-\delta_g^2/2\sigma^2}\label{eq:density1}&\\ 
&P_{LN}(\delta_g) = \frac{1}{\sqrt{2\pi}\sigma}e^{-(ln(1+\delta_g)+\tilde{\sigma}^2/2)^2/2\tilde{\sigma}^2}/(1+\delta_g)\label{eq:density2}&
\end{eqnarray}
\begin{eqnarray}
P_{E}(\delta_g) =& \frac{1}{\sqrt{2\pi}\sigma}e^{-\delta_g^2/2\sigma^2}\big( 1+\frac{\sigma S_3}{6}H_3(\delta_g / \sigma) \nonumber \\
&+ \frac{\sigma^2S_4}{24}H_4(\delta_g / \sigma)+ \frac{10\sigma^2S_3^2}{720}H_6(\delta_g / \sigma) \big),
\label{eq:density3}
\end{eqnarray}
where $\sigma^2 = \overline{\xi}$, $\tilde{\sigma}^2 = \sqrt(1+\overline{\xi})$ and $H_i$ is the $i$th Hankel-function. The above approximations are plotted in Figure \ref{fig9} for galaxy counts in cubes with a linear size of $100~h^{-1}$Mpc from the spectroscopic LRG sample. The best match around the tails is the Edgeworth expansion, while around the mean all of these approximations provide qualitatively similar results. The left panel of Figure \ref{fig9} shows the underlying continuous galaxy PDFs, the deconvolved CIC distribution, while the right shows them convolved with the Poisson-distribution and the galaxy counts. 

To estimate the density of the 50 supervoids and 50 superclusters of a certain size we generated $V_{survey}/V_{cell}$ random numbers with the PDFs given above, where $V_{survey}$ refers to the volume of the survey and $V_{cell}$ refers to the volume of a superstructure. This approximation slightly overestimates the number of independent cells. Then we stored the lowest and highest 50 of the numbers. Repeating this several times gives the distribution of the extrema. This method, however, is too slow for calculating a covariance matrix from hundreds of simulations. Since we only deal with linear ISW here, as we explain it in the next Subsection, we need only the mean of the extrema rather than their whole distribution. A satisfactory approximation for the mean in the case of voids is:
\begin{eqnarray}
&\overline{\delta_g} = \int_{-\infty}^{\delta_{max}}  P(\delta_g)\delta_g d \delta_g ,&
\label{eq:delta}
\end{eqnarray}
where $\delta_{max}$ is given by $50V_{cell}/V_{survey} = \int_{-\infty}^{\delta_{max}}  P(\delta_g)d \delta_g$.
For clusters the limits of the integration change to $\int_{\delta_{max}}^{\infty}$. In our tests we found the difference to be about a few percent. As we use the linear bias model, a further division by the bias yields $\delta$.

   \subsection{The Profile of Supervoids and Superclusters and the ISW Signal}
   
       \subsubsection{The Average Profile of Superstructures}
       
In general the ISW effect is determined by an integral along the path of a CMB photon \citep{SW1967}:
\begin{eqnarray}
\frac{\Delta T}{T} = -\frac{2}{c^2}\int d \tau \frac{\partial \Phi (r(\tau),\tau) }{\partial \tau},
\label{eq:isw}
\end{eqnarray}
where $\tau$ denotes the conformal time.

According to \cite{Rudnick2007} a simple estimate of the linear ISW effect, an underdense or an overdense spherical region at redshift $z$ causes a
\begin{eqnarray}
&\frac{\Delta T}{T} \approx \Omega_m \big( \frac{r_c}{c/H_0} \big)^3(1+2z)(1+z)^{-2}\delta
\label{eq:Rud}&
\end{eqnarray}   
temperature shift at its center on the CMB, where $r_c$ is the comoving radius of the sphere. In the derivation the authors approximated the potential with a top hat which implies a compensated void or cluster. Here, instead, we propose a profile motivated by Gaussian statistics. For a spherically symmetric object the Newtonian potential can be calculated easily as:
\begin{eqnarray}
&\Phi (r) = -\frac{3\Omega_m}{8\pi}\bigg( \frac{H_0}{c} \bigg)^2 \int_{r}^{\infty} \frac{M(\tilde{r})}{\tilde{r}^2}d \tilde{r},
\label{eq:pot}&
\end{eqnarray}
where $M(r) = 4\pi \int_{0}^{r}d \tilde{r} \tilde{r}^2\delta (\tilde{r}) $.
The density contrast at distance $r$ from the center can be obtained with the condition that the average density inside $r_c$ is known:
\begin{eqnarray}
& P(\delta_{in}) = \frac{1}{\sqrt{2\pi \big<\delta_{in}^2\big>_{uc}}}\exp \bigg( -\frac{\delta_{in}^2}{2\big<\delta_{in}^2\big>_{uc}} \bigg), & \\
& P(\delta (r),\delta_{in}) = \frac{1}{\sqrt{2\pi |C|}}\exp \bigg( -\frac{1}{2}\vec{\delta} C^{-1}\vec{\delta} \bigg), & \\
& P(\delta (r)|\delta_{in}) = P(\delta (r),\delta_{in})/P(\delta_{in}) &
\end{eqnarray}
where $\big<...\big>_{uc}$ refers to unconditional ensemble averaging and $\delta_{in}$ is the average density measured inside $r_c$. For short we use $\vec{\delta} = (\delta (r),\delta_{in})$ and $C = \big<\vec{\delta} \otimes \vec{\delta}\big>_{uc}$. In practice $\big<...\big>_{uc}$ can be computed with Monte Carlo simulations robustly. From these the expected density is
\begin{eqnarray}
& \big<\delta (r)\big> = \frac{\big<\delta (r) \delta_{in}\big>_{uc}}{\big<\delta_{in}^2\big>_{uc}}\delta_{in}.\label{eq:prof} &
\end{eqnarray}
This profile is not compensated inside a finite radius.

       \subsubsection{The Uncertainty of the Profile and the ISW Effect}
            
It is useful to calculate the uncertainty of the profile in order to get an estimate of the uncertainty of the potential and the ISW effect. We would like to point out that for a correct treatment of the potential one should drop the assumption of spherical symmetry. We chose to optimize the accuracy and speed by keeping the spherical approximation. We use the Gaussian model as before:
\begin{eqnarray}
& P(\delta(r_1),\delta(r_2),\delta_{in}) = \frac{1}{\sqrt{2\pi |C_3|}}\exp \bigg( -\frac{1}{2}\vec{\delta}_3 C_3^{-1}\vec{\delta}_3 \bigg) & , \\
& P(\delta(r_1),\delta(r_1)|\delta_{in}) = P(\delta(r_1),\delta(r_1),\delta_{in})/P(\delta_{in}), &
\end{eqnarray}
where $\vec{\delta}_3 = (\delta (r_1),\delta (r_2),\delta_{in})$ and $C_3 = \big<\vec{\delta}_3 \otimes \vec{\delta}_3\big>_{uc}$. From this the covariance between shells at $r_1$ and $r_2$ is:
\begin{eqnarray}
Cov(r(1),r(2)) = & \big<\delta(r_1) \delta(r_2)\big> - \big<\delta(r_1)\big>\big<\delta(r_2)\big> \nonumber \\
 = & \frac{1}{\big<\delta_{in}^2\big>_{uc}}\bigg[ \big<\delta(r_1) \delta(r_2)\big>_{uc}\big<\delta_{in}^2\big>_{uc} \nonumber \\ 
& -\big<\delta(r_1) \delta_{in}\big>_{uc}\big<\delta(r_2) \delta_{in}\big>_{uc}\bigg] .
\end{eqnarray}  

\begin{figure}
        \begin{center}	
        \includegraphics[scale=0.45]{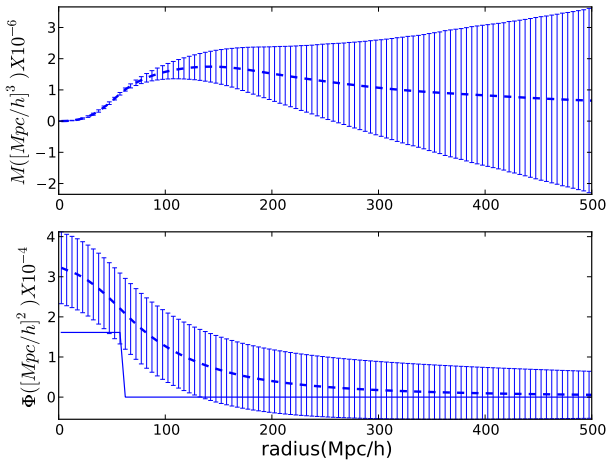}
        \end{center}
        \caption{On the upper panel $M(r) = 4\pi \int_{0}^{r}d r r^2\delta (r) $ is potted for a supercluster with $\delta_{in} = 1$ for $r_c = 62~h^{-1}$Mpc. On the lower panel we plot the potential (dashed line) and the top hat potential (see the text for details).\label{fig11}}
\end{figure}

On the upper panel of Figure \ref{fig11} we plot $M(r)$ for $r_c = 100/(4\pi /3 )^{1/3}\mpc$. The curve is normalized so that $M(r_c) = \frac{4\pi r_c^3}{3}$. The error bars are:
\begin{eqnarray}
&\Delta M(r) = \sqrt{(4\pi)^2 \int_0^r \int_0^r Cov(r_1,r_2) r_1^2 r_2^2 d r_1 d r_2}  &.
\end{eqnarray}
On the lower panel of Figure \ref{fig11} we plot the potential of such an object as in Equation (\ref{eq:pot}) along with the top hat potential from \cite{Rudnick2007}.

\begin{figure}	
        \begin{center}
        \includegraphics[scale=0.45]{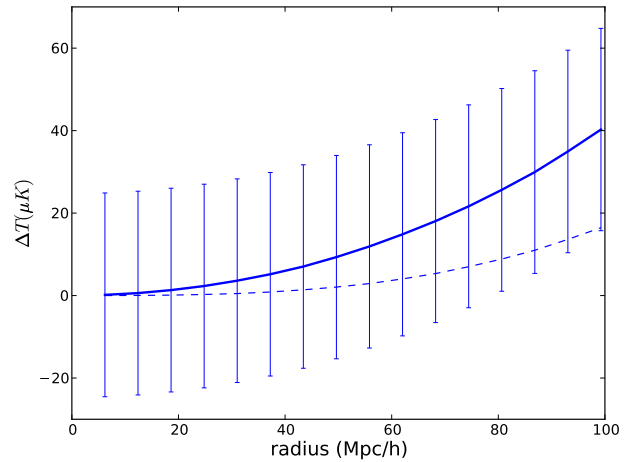}
        \end{center}
        \caption{$\Delta T$ for a photon traveling through the center of a supervoid at redshift 0.52 with $\delta_{in} = 1$ against its radius is plotted (solid line). The dashed line comes from Equation (\ref{eq:Rud}), an approximation using compensated profile.\label{fig12}}
\end{figure}

Since the potential is only Equation (\ref{eq:pot}) and in linear theory its time dependence is relatively simple $\Phi(r,\tau) = \Phi(r) \frac{D(\tau)}{a(\tau)}$, it is straightforward to integrate Equation (\ref{eq:isw}) numerically. In our calculations we placed the density fluctuation at the median redshift $(z=0.53)$ of the SDSS photometric LRG survey so that we can compare our result to actual measurements \citep{Granett2008}. In Figure \ref{fig12} we plot $\Delta T$ versus the comoving radius ($r_c$) of the superstructure, we also plot $\Delta T$ according to Equation (\ref{eq:Rud}). Here, we used $\delta_{in} = 1$. It is clear that Equation (\ref{eq:Rud}) underestimates the ISW effect.

In Figure \ref{fig13}, we used Equation (\ref{eq:delta}) to calculate $\delta_{in}$. The PDF was measured in an SDSS LRG subsample at $z=0.28$ median redshift (see \ref{data} for details) and scaled to the subsample described in \cite{Granett2008}, which is located at $z=0.53$. This means the scaling of the variance, $S_3$ and $S_4$ according to linear dynamics. The value for $V_{survey}$ came from the properties of the survey in \cite{Granett2008}. All three assumptions in Equations (\ref{eq:density1}--\ref{eq:density3}) about the density distribution give similar results. We also plot an estimate based on raw data without deconvolution. The dashed line is the ISW effect according to Equation (\ref{eq:Rud}). We plotted both the supervoids and superclusters. The intrinsic fluctuations of the matter density (see Figure \ref{fig12}) and the uncertainty of the tail of the PDF add up. On the horizontal axis the scale is the linear size of the cell we measured CIC in.

\section{Discussion}
\label{dis}
We measure the first few moments of the matter PDF and give an estimate of the linear ISW effect owing to the largest density fluctuations on $100~h^{-1}$Mpc scales. 

\begin{figure*}
        \begin{center}	
        \includegraphics[scale=0.6]{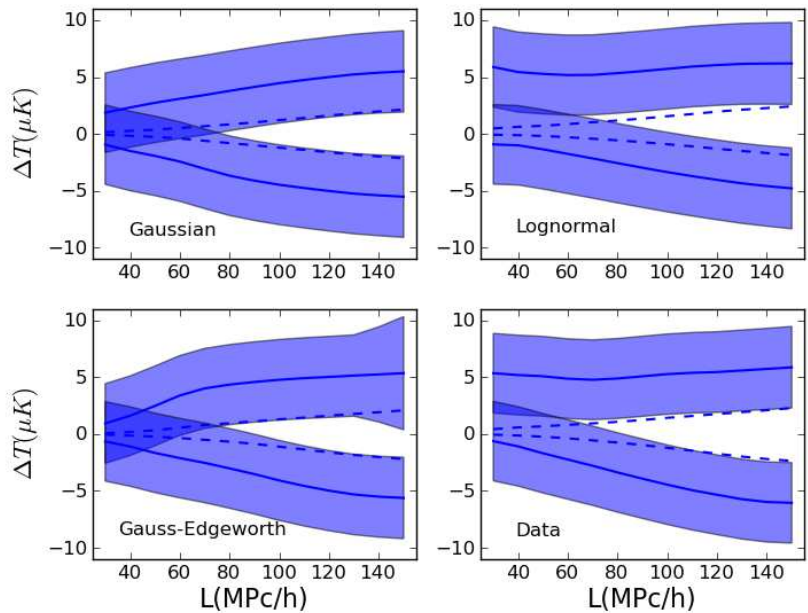}
        \end{center}
        \caption{Expected ISW effect from the average of 50 supervoids and superclusters from a survey similar to the SDSS DR4 photometric LRG sample. The dashed line is Equation (\ref{eq:Rud}). On each panel we use a different approximation of the matter PDF.\label{fig13}}
\end{figure*}

First, we measure CIC in a volume-limited subsample of the SDSS DR7 spectroscopic LRG data. We compare the variance, $S_3$ and $S_4$ to their values in $\Lambda$CDM. Despite that we use the lowest order approximation we find agreement between data and prediction (Figure \ref{fig6-8}). However, it cannot be excluded that more complex models can fit the data better especially on nonlinear scales. As can be seen from Figure \ref{fig6-8}, $S_3$ and $S_4$ differ slightly from the lowest order predictions possibly due to nonlinearities, although this difference is not statistically significant. We approximate the tail of the matter PDF with analytic functions to get the density extrema. We calculate the expected radial profile of supervoids and superclusters with the condition that their average density inside a sphere is known. We estimate the average linear ISW signal of 50 of the most significant from each in a realistic survey. As can be seen from Figure \ref{fig13}, it is plausible that linear ISW can produce the results presented in \cite{Granett2008}. They used a compensating top hat filter with inner radius of $4^{\circ}$ to get $7.9\pm3.1\mu K$ for clusters and $-11.3\pm3.1\mu K$ for voids. In comparison, our estimates for the temperature at the center of the same stacks from Figure \ref{fig13} are $5.5\pm3.5$ and $-5.5\pm3.5$ in the case of a Gaussian PDF. We plot the linear size of a cube on the $x$-axis. The projection of a sphere with the same volume gives the corresponding angle. In case of L = $150~h^{-1}$Mpc, this is $3^{\circ}.8$. While our errors originate from the fluctuations of the ISW signal, theirs come from the primary CMB anisotropies. Thus they are independent and our calculation is fully consistent with \cite{Granett2008}. Our estimate is robust. It is not affected significantly by the details of the matter PDF. The error bars can be tightened if the volume of the survey is larger and if more images are stacked. The former would reduce the cosmic error on the scales we study, while the latter would give a more accurate measurement of the average profile. 

We take one step toward cosmological parameter estimation with calculating the expected linear ISW signature of supervoids and superclusters. The next step can be to depart from the spherically symmetric model that we use for the sake of simplicity. Anisotropic fluctuations in the matter density around the center of a superstructure might give a quantitatively different error estimate. We also ignore any nonlinearities. We work with linear scales but we also probe the highest and lowest densities. The latter calls for a biasing model more complex than linear. We also ignore the nonlinear ISW, the Rees--Sciama effect \citep{RS1968}. It has been shown that it is small compared to the linear part at low redshifts (see \citealt{Cai2009,Cai2010}). Another possible improvement is to use general relativity (GR) instead of Newtonian. A model of compensated voids based on GR is discussed in \cite{Inoue2007} and \cite{Inoue2010}. We also ignore the correlation between the objects.

Our result suggests that the void needed to produce detectable anomalies on the CMB is smaller than previously estimated. The CMB Cold Spot \citep{Cruz2005} has been considered consistent with a compensated void having $\delta = -0.3$ and a radius of $200~h^{-1}$Mpc by \cite{Sakai2008} in agreement with the heuristic argument of \cite{Rudnick2007}. From Figure\ \ref{fig12} one can see that a void of similar size needs to have a much less significant underdensity in our Gaussian, non-compensated model. The signal from a top-hat potential is less than half of the signal of a realistic void. This makes the detection of such voids harder in today's sparse catalogs (see \citealt{Granett2010} and \cite{Bremer2010}).

We thank Ben Granett for his useful comments. The authors were supported by NASA grants NNX10AD53G and NNG06GE71G.

\bibliographystyle{apsrmp}
\bibliography{ms}

\end{document}